\documentclass[%
reprint,
 amsmath,amssymb,
 aps,
]{revtex4-2}

\usepackage{graphicx}%
\usepackage{svg}
\usepackage{dcolumn}%
\usepackage{bm}%

\begin{document}

\preprint{APS/123-QED}

\title{The Persistent Clock of Turbulent Thermal Convection}

\author{Lázaro Martínez-Ortíz}
\author{Youri H. Lemm}
\author{Herman J. H. Clercx}
\author{Rudie P. J. Kunnen}
 \email{r.p.j.kunnen@tue.nl}

\affiliation{Fluids and Flows Group, Department of Applied Physics and Science Education and J. M. Burgers Center for Fluid Dynamics, Eindhoven University of Technology, P.O. Box 513, 5600 MB Eindhoven, The Netherlands}%

\date{\today}%

\begin{abstract}

The large-scale circulation (LSC) of turbulent convection is a prominent feature of its dynamics and forms the basis for descriptive theories. We show, using experimental and numerical results from thermal convection in a cylindrical cell, that the LSC possesses a persistent internal `clock': its pulsating velocity as a function of time is described by a constant value of the parameter $U/(lf)$ where $U$ is the mean velocity, $f$ the pulsation frequency, and $l$ the characteristic length scale. By introducing a narrow sidewall barrier, we can trip the LSC, forming a pair of interconnected rolls stacked above and below the barrier. They independently exhibit the same value for the ratio $U/(lf)$, even for vertically asymmetric pairs, indicating signs of synchrony. Thus, this parameter establishes a direct connection between plume-shedding dynamics and the flow topology.

\end{abstract}

\maketitle

For over a century, a fluid layer heated from below and cooled from above has served as the established model of buoyancy-driven convection, commonly known as Rayleigh-Bénard convection (RBC) \cite{rayleigh1916lix,benard1900tourbillons}. A hallmark of turbulent RBC in confined geometries is the emergence of a large-scale circulation (LSC) \cite{ahlers2009heat}, with hot fluid rising on one wall and cold fluid sinking on the opposite, organizing into  a single cell-spanning roll. The LSC shows a remarkable rich dynamical variability \cite{villermaux1995memory,brown2005reorientation,sun2005azimuthal,brown2006rotations}, while its large-scale structure persists \cite{tsuji2005mean,wei2021persistence,xie2017turbulent,wagner2015heat}. This resilience provides the foundation for phenomenological models \cite{grossmann2000scaling,ahlers2009heat,stevens2013unifying}, which explicitly rely on the presence and structure of the LSC to make their predictions \cite{ahlers2009heat,villermaux1995memory,vanderpoel2011connecting,schindler2022collapse,lohse2024ultimate,xu2025restoration}. However, early studies employing strong obstructions to disrupt the LSC \cite{ciliberto1996large,xia1997turbulent,xia1999turbulent} challenged these theoretical expectations, showing that heat transport was largely insensitive to flow structure.

The LSC arises from the collective organization of hot and cold thermal plumes, which tend to cluster and move synchronously, producing periodic near-wall temperature and velocity signals \cite{qiu2001onset, villermaux1995memory}. Plume circulation and mean wind have been proposed to be mutually slaved \cite{cioni1997strongly,sun2005three}, but the precise nature of this coupling remains unclear, whether the LSC drives the motion of the plumes or is itself sustained by their buoyant driving
remains an open question.

In this Letter, we address the problem by directly controlling the flow topology. We show that a small perturbation (a narrow ring-shaped sidewall barrier at mid-height) suffices to reorganize the LSC. At low forcing, it remains stable, but at higher driving it splits into a pair of interconnected counter-rotating circulations with stepwise boundary-layer thinning. In this new topology, the characteristic velocity retains its original scaling, while the pulsation frequency of the flow doubles. A universal relation emerges, linking the circulation timescale to the plume-shedding frequency. The pair of rolls are dynamically linked, with frequencies connected through the same relation that persists even when they are asymmetric. These results expose a direct quantitative connection between plume dynamics and mean flow and opening new avenues for using RBC to probe the interplay between circulation structure, synchrony, and turbulence.

\begin{figure}[ht]
\centering
\includegraphics[width=0.48\textwidth]{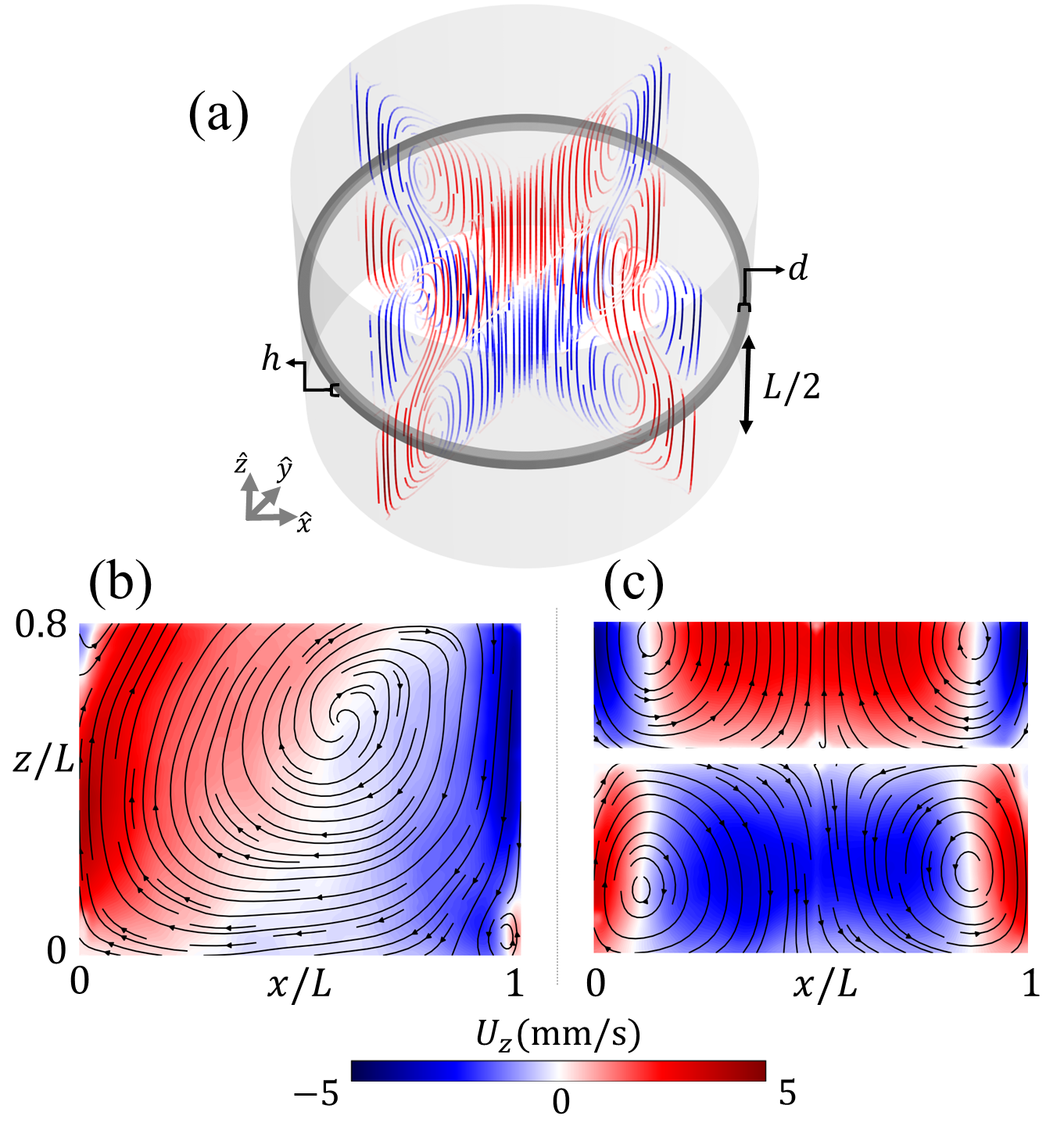}
\caption{\label{f1}(a) Sketch of the $\Gamma = 1$ cylindrical cell, where a narrow ring-shaped barrier is mounted at mid-height along the sidewall; it can be removed or repositioned. Oblique views of streamlines from two perpendicular cross-sections reveal the nearly axisymmetric flow structures inside. (b), (c) Measurements of the mean vertical velocity $U_z$ overlaid with 2D streamlines. The field of view covers about $80\%$ of the full height $L$. (b) For a smooth sidewall, the LSC dominates. (c) With the barrier, the LSC is replaced by a four-roll structure within the measurement field; the barrier appears as a white band where it obstructs a small portion of the field. (a) Data from numerical simulations; (b,c) experiments, all at $\mathrm{Ra}=1.03\times10^9$.}
\end{figure}

Experiments are carried out on a cylindrical sample filled with water at a mean temperature $\overline{T}\approx30$ $^\circ\mathrm{C}$. The cell has diameter and height $D = L = 0.2\,\mathrm{m}$ (aspect ratio $\Gamma = 1$), with copper top and bottom plates and a Plexiglas sidewall. A ring-shaped Plexiglas barrier ($d = 10$ mm wide and $h = 8$ mm thick) is mounted around the sidewall (see Fig. \ref{f1}(a)), positioned at heights $L/2$, $L/3$ or $L/4$. The convective flow is described by the Rayleigh number $\mathrm{Ra} = \alpha g \Delta L^3 / (\nu \kappa)$, which quantifies the balance between buoyancy and dissipative effects, and the Prandtl number $\mathrm{Pr} = \nu / \kappa$, the ratio of momentum to thermal diffusivity. Here, $\alpha$ is the thermal expansion coefficient, $g$ gravity, $\Delta$ the imposed temperature difference, $\nu$ the kinematic viscosity, and $\kappa$ the thermal diffusivity. The experiments cover the range $3.8 \times 10^8 \leq \mathrm{Ra} \leq 4.7 \times 10^9$, with an average Prandtl number of $\mathrm{Pr} \approx 5.8$.

The flow field is measured using Particle Image Velocimetry (PIV) \cite{raffel2018piv}. To reduce optical distortions caused by the cylindrical sidewall, a square Plexiglas jacket filled with water is placed around the cell. The PIV system employs a vertically aligned laser sheet of $2$ mm thickness, positioned at the mid-plane of the cell, $y=L/2$, and a high-resolution camera ($2560 \times 2048$ pixels) that captures the motion of $20$-$\mu$m-diameter polyamide tracer particles at $15$ frames per second. Each velocity field is reconstructed on a $79 \times 63$ vector grid, yielding a spatial resolution of $2.4$ mm.

To complement the experiments and gain access to the full three-dimensional velocity and temperature fields, we conduct DNS of the incompressible Navier-Stokes and heat equations for a Boussinesq fluid. The formulation is nondimensionalized using the temperature difference $\Delta$, height $L$, and the free-fall velocity $U_{ff}=\sqrt{\alpha g\Delta L}$. No-slip velocity conditions are applied on   all boundaries, the sidewalls are thermally insulating, while the top and bottom plates are held at fixed temperatures. The equations are discretized on a cylindrical grid using a second-order finite-difference scheme with third-order Runge-Kutta time stepping. Further details on the numerical scheme can be found in Refs. \cite{verzicco1996cylindrical,verzicco2003RBC}. The simulations are performed with parameters that match those of the experiments for three Ra values: $3.8\times 10^{8}$, $1.03\times 10^{9}$, and $5.0\times 10^{9}$. To implement the sidewall barrier numerically, we employ the volume penalization method \cite{engels2015numerical}: we introduce an additional term in the Navier–Stokes equations, active only within the barrier region, forcing the velocity towards zero there and thus mimicking an impermeable boundary. Heat diffusion is solved in the barrier region in the same way as in the fluid, using the same thermal diffusion coefficient (see the Supplemental Material for additional information).

Figure \ref{f1}(b,c) shows the mean velocity field under identical experimental conditions ($\mathrm{Ra} = 1.03 \times 10^9$), the only difference being the presence or absence of the barrier. In the smooth-wall configuration, Fig. \ref{f1}(b), the flow is dominated by a single large-scale convective roll, with upwelling and downwelling on opposing sides of the cell (the LSC), two smaller counterrotating secondary rolls appear near diagonally opposite corners, consistent with the expected structure in this regime. The LSC spans the entire length of the container, with a characteristic scale comparable to the size of the system $L$. Alternatively, when a barrier is introduced at mid-height ($L/2$), the flow reorganizes dramatically, Fig. \ref{f1}(c). The classical single-roll LSC structure is disrupted, and the velocity field exhibits a four-roll structure in the vertical mid-plane. The location of the barrier sets the scale of the emerging flow structures by splitting the domain. Rather than a single full-length circulation, two half-sized pairs of counter rotating vortices form above and below the barrier, extending from the horizontal top and bottom plates to the sidewall obstruction at $L/2$. Due to experimental constraints, the measurements do not cover the full vertical extent of the cell. About $20\%$ of the total cross-sectional area near the top plate remains obscured. However, the symmetries of the problem allow us to draw robust conclusions from the accessible portion. Figure \ref{f1}(a) shows that the modified flow is nearly axisymmetric, forming an interconnected pair of donut-like, counter-rotating vortices above and below the barrier. These pairs interconnect to produce figure-eight–shaped structures in every vertical plane, linked in a complex three-dimensional pattern. We refer to the top (bottom) vortex pair as a Half-Cell Circulation (HCC).

\begin{figure}[ht]
\centering
\includegraphics[width=0.45\textwidth]{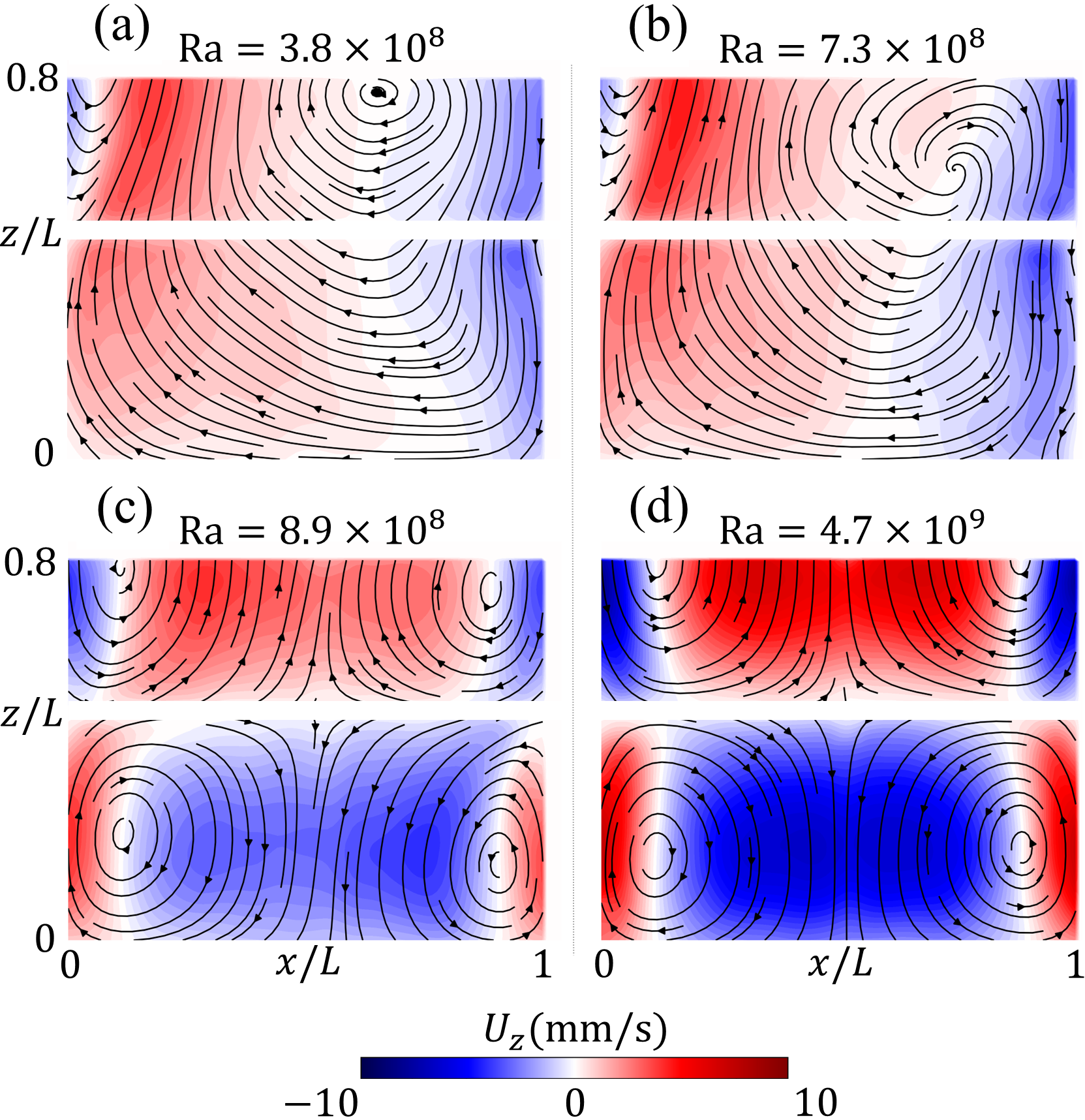}
\caption{\label{f2}Flow topology at different driving strengths ($\mathrm{Ra}$), shown through mean vertical velocity ($U_z$) fields overlaid with 2D streamlines for the barrier case.
(a,b) At lower thermal driving, the LSC persists as the dominant flow component.
(c) As driving increases the flow breaks into four rolls in the vertical plane.
(d) The HCC remains robust even at higher levels of turbulence.}
\end{figure}

As a step in further exploring the underlying causes of the observed changes in flow structure, we measured velocity fields at various intensities of thermal forcing. Figure \ref{f2} highlights the flow topology for different Ra through 2D streamlines. Notably, the LSC remains the dominant flow feature at low thermal driving, $\mathrm{Ra}=3.8 \times 10^8$ and $7.3 \times 10^8$ (Fig. \ref{f2}(a,b)), circulating over the barrier as if it were unobtrusive. As Ra increases, a transition to the HCC is observed after $\mathrm{Ra}\approx 8 \times 10^8$, Fig. \ref{f2}(c). This new flow topology persists as the turbulence intensifies, observed up to Ra $= 4.7 \times 10^9$, Fig. \ref{f2}(d). This shows that the LSC can be disrupted, but only when the barrier length becomes a noticeable perturbation to the system; below this threshold, the LSC structure persists despite the obstacle.

The origin of this transition is examined  by focusing on the viscous boundary layers (BLs), which are central to momentum and heat transport \cite{grossmann2000scaling, ahlers2009heat}. Near the boundaries, the time-averaged wall-parallel velocity component $U_{\parallel} = \langle u_{\parallel}(t) \rangle$ shows the classical Prandtl–Blasius (PB) structure: linear growth from the wall, a peak $U_{\max}$ at distance $\delta_u$, and a gradual decay toward bulk values \cite{prandtl1905uber, blasius1907grenzschichten, ahlers2009heat, wagner2012boundary,schlichting2016boundary}. Building on this foundation, Grossmann-Lohse (GL) theory \cite{grossmann2000scaling, ahlers2009heat} offers a unified description of turbulent thermal convection by treating the bulk and boundary layers (BLs) separately. A key assumption is that the kinetic BL obeys Blasius-type scaling, with $\delta_u/L \propto \mathrm{Re}^{-1/2}$, where $\mathrm{Re} = U_{\mathrm{max}} L / \nu$ is the Reynolds number. Consistent with this, experiments support a power-law dependence of $\delta_u$ on thermal forcing, with $\delta_u/L\propto$ Ra$^{-0.27 \pm 0.01}$ \cite{sun2008experimental, ahlers2009heat}. This scaling implies a progressive thinning of the BL with increasing thermal forcing and has been validated over a wide range of parameters \cite{sun2008experimental, tai2021heat}.

\begin{figure}[ht]
\centering
\includegraphics[width=0.45\textwidth]{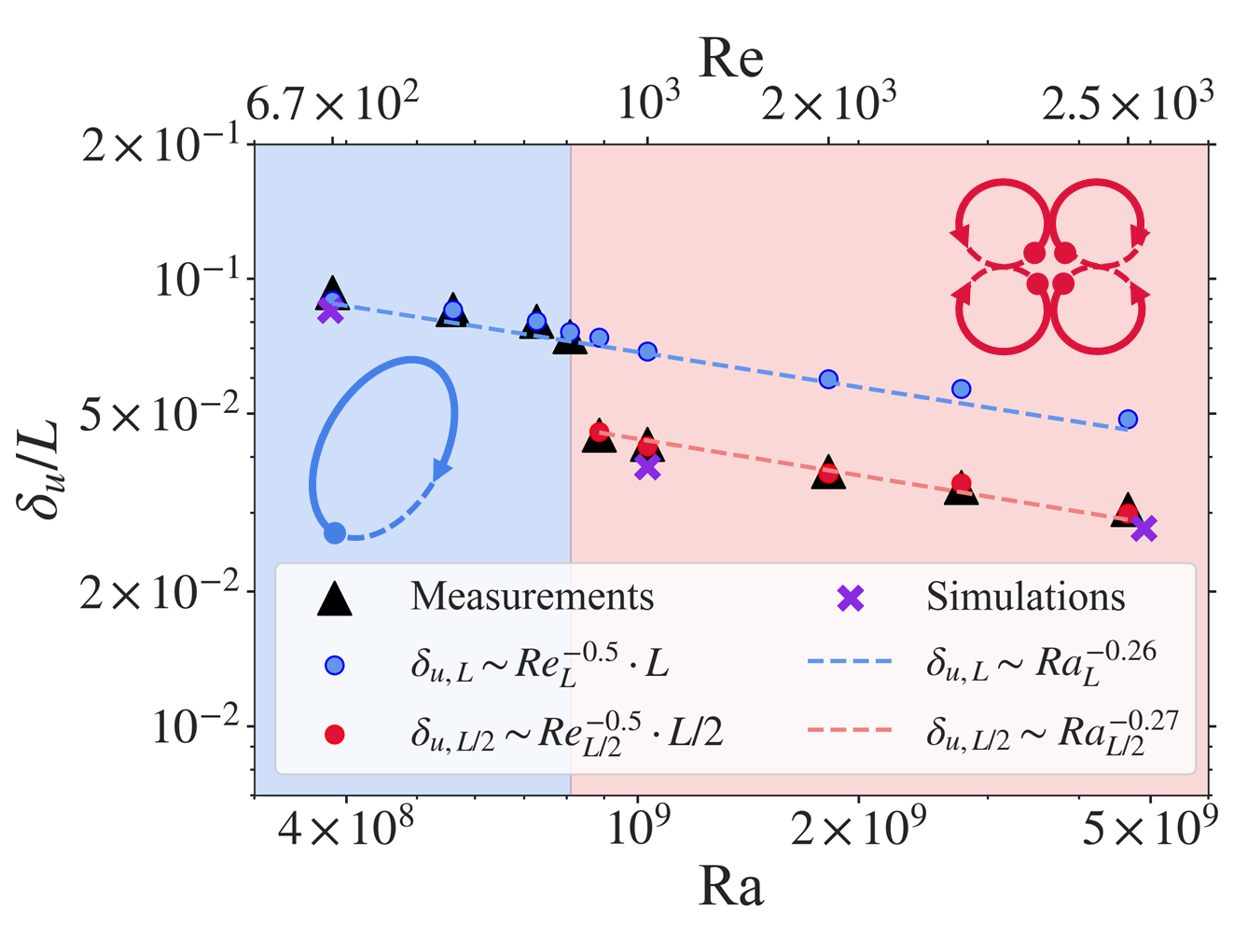}
\caption{\label{f3} Normalized BL thickness as a function of $\mathrm{Ra}$ and $\mathrm{Re}$.
In the blue-shaded region we observe a single LSC; in the pink-shaded region we find two stacked HCCs. Black triangles mark direct measurements from the velocity profiles, using the maximum-velocity position relative to the wall. Circular markers indicate the Reynolds-number scaling laws using the measured $U_{max}$; blue points use the classical definition $\mathrm{Re}_L$,  red circles use the rescaled $\mathrm{Re}_{L/2}$. Dashed lines represent the scaling laws based on $\mathrm{Ra}_L$ and $\mathrm{Ra}_{L/2}$.
} 
\end{figure}

The transition is then triggered when the viscous boundary layers thin to the point of being comparable to the barrier thickness. Figure \ref{f3} shows our measured $\delta_u$ as a function of Ra and Re. At low forcing ($\mathrm{Ra}\lesssim 8\times 10^8$), the measurement and simulation results align with the classical scaling laws. For $\mathrm{Ra}\gtrsim 8\times 10^8$, where the flow structure transition has occurred, the measurements exhibit a step-like deviation from the classical scaling behavior. The BL becomes thinner than predicted by the PB and GL theories. We attribute this deviation to the observed change in the flow structure, from a single large-scale roll to two half-length structures, effectively reducing the characteristic length scale from $L$ (blue-shaded region on the left of Fig. \ref{f3}) to $L/2$ (pink-shaded region on the right of Fig. \ref{f3}). To account for this change, the dimensionless numbers are rescaled using the reduced effective length, such that $\mathrm{Re}_{L/2} = U_{\mathrm{max}} (L/2)/\nu$ and $\mathrm{Ra}_{L/2}  = \alpha g \Delta (L/2)^3/(\nu\kappa)$, while $\mathrm{Re}_L$ and $\mathrm{Ra}_L$ retain their standard definitions. With these modifications, the measurements realign with the classical scaling laws and recover comparable prefactors. Specifically, when the LSC remains the dominant flow component, we find $\delta_{u,L} = (15.0 \pm 1.0)\mathrm{Ra}_L^{-0.26 \pm 0.01} L$ and $\delta_{u,L} = (2.3 \pm 0.1)\mathrm{Re}_L^{-0.50 \pm 0.01} L$, as indicated by the blue dashed line and circles in Fig. \ref{f3}. After the transition, a slightly steeper scaling is recovered for $\mathrm{Ra}$, $\delta_{u,L/2} = (13.5 \pm 1.0)\mathrm{Ra}_{L/2}^{-0.27 \pm 0.01} L/2$, which remains consistent within the statistical error, and for $\mathrm{Re}$, $\delta_{u,L/2} = (2.1 \pm 0.1)\mathrm{Re}_{L/2}^{-0.50 \pm 0.01} L/2$, as shown by the red dashed lines and circles in Fig. \ref{f3}. 

\begin{figure*}[ht]
\centering
\includegraphics[width=\textwidth]{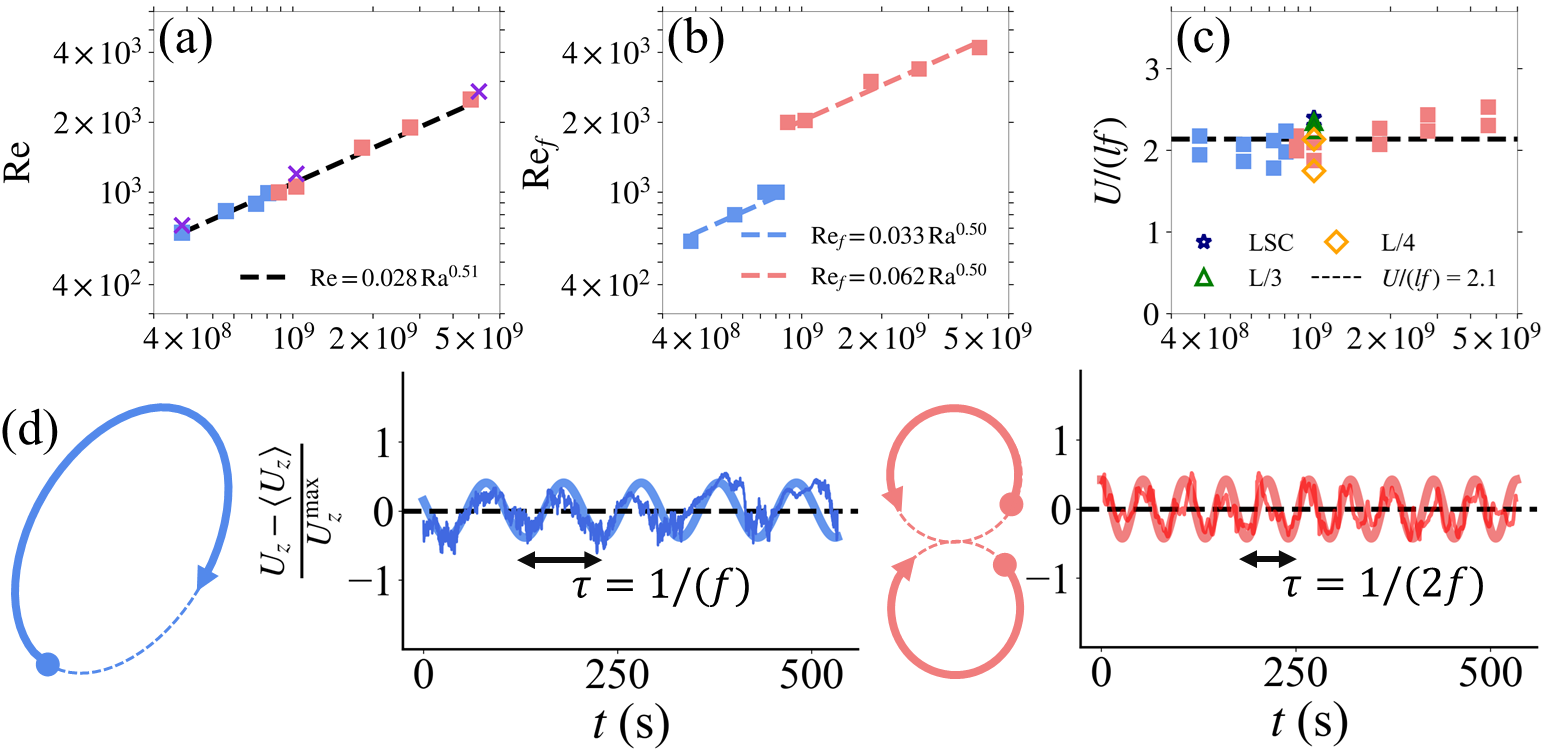}
\caption{\label{f31}  (a) $\mathrm{Re}$ as function of $\mathrm{Ra}$ for the barrier case. Squares denote experimental measurements (blue for LSC, pink for HCC), purple crosses indicate DNS results, the dashed line is a fit. (b) $\mathrm{Re}_f$ as function of $\mathrm{Ra}$ for the barrier case (dashed blue and pink lines represent the fits corresponding to the one-roll and two-pairs states). (c) Dependence of $U/lf$ on $\mathrm{Ra}$. The two points per $\mathrm{Ra}$ correspond to measurements taken at the left/right sides when the LSC is dominant and at the top/bottom rolls after the transition (see Supplemental Material Fig. S1). The horizontal discontinuous line represents the fitted value $c=2.1$. Open symbols for $\mathrm{Ra}=1.03\times10^9$ correspond to cases where the barrier is not at $L/2$: LSC (without the barrier), $L/3$, and $L/4$. (d) Time series of near-wall velocity at $\mathrm{Ra}=1.03\times10^9$ in non-barrier and barrier cases (dark-shaded curves). After the transition, the flow pulsates at a higher frequency. The light-shaded lines correspond to the dominant frequency component of the signal’s Fourier decomposition.
}
\end{figure*}

So far, our arguments have relied solely on the shift in the characteristic system scale. Examining the Reynolds–Rayleigh dependence (without applying any additional rescaling) provides further support for recovering the scaling laws using the new scale ($L/2$). Figure \ref{f31}(a) shows this relationship, where blue squares represent the flow before the transition and pink squares correspond to the flow after the transition. A power-law, $\mathrm{Re}_L = (0.028 \pm 0.003)\mathrm{Ra}_L^{0.51 \pm 0.02}$, fits the data, in agreement with classical scalings \cite{ahlers2009heat, sun2008experimental}. This indicates that the characteristic velocity in the flow remains $U_{\max}$, as in the case without barriers, so the only significant change in the system is in the length scale.

The Reynolds number can also be characterized using the pulsation frequency $f$. This frequency can be extracted from near-wall temperature or velocity time series, where a periodic signal arises due to the coupling between boundary-layer instabilities (leading to the release of thermal plumes) and the slow convective motion of the LSC. In this scenario, the plumes and the LSC are mutually slaved \cite{villermaux1995memory,cioni1997strongly,sun2005three,Brown2007}. Using this connection, one can define a frequency-based Reynolds number, $\mathrm{Re}_f = c f L^2 / \nu$, which has been reported to exhibit values and scaling similar to the velocity-based Reynolds number with scaling constant $c=2$, though $c(\mathrm{Ra})$ may rise beyond $2$ for larger $\mathrm{Ra}$ \cite{sun2005scaling,Brown2007,ahlers2009heat}.

 In our case, after the flow structure transition, we observe a surprising effect: $f$ is twice as large than expected, causing a step increase in $\mathrm{Re}_f$, Fig. \ref{f31}(b,d). This behavior can also be understood from the reduction of the effective length scale in the system. Since the characteristic velocity can be expressed as $U_\mathrm{max} = c(\mathrm{Ra}) L f\approx 2 L f$, halving the length of the circulation path to $L/2$ requires doubling $f$ to maintain the same velocity. This suggests that the flow adapts, pulsating at a higher frequency after its scale is reduced. Based on this observation and considering that previous studies have shown that the two definitions of the Reynolds number closely coincide over a wide range of parameters ($\mathrm{Re}_L = \mathrm{Re}_f$) \cite{Brown2007,xia2007two, sun2005scaling,ahlers2009heat}, a conserved quantity can be identified, $U_i/(l_i f_i) = \text{const}$, where $i$ denotes a flow structure of characteristic size $l_i$ and velocity $U_i$. In Fig. \ref{f31}(c), we show the $\mathrm{Ra}$ dependence of this ratio, which remains nearly constant at $c=2.1$, displaying only a weak growth at higher turbulence levels. Physically, we argue that the plume-shedding frequency has a characteristic relationship with the LSC time scale $\tau=1/f$, or equivalently to the shear rate $\dot{\gamma}=U_i/l_i$, so $\tau_i\dot{\gamma}_i=U_i/(l_i f_i)=\text{const}$. In other words, the ``clock of circulation'' persists according to the conserved dimensionless parameter $U_i/(l_i f_i)$ even when the flow topology undergoes significant changes. The weak upward trend in Fig. \ref{f31}(c) is attributed to the reported increase in the effective circulation path with $\mathrm{Ra}$ (i.e., $c(\mathrm{Ra})$ growing beyond $2$) \cite{sun2005scaling}.
 Interestingly, this parameter is similar to the Strouhal number ($St^{-1}=U/Lf$), which, in other contexts of oscillating flows, has also been observed to remain approximately constant over a wide range of Reynolds numbers \cite{lienhard1966synopsis}.  

\begin{figure}[ht]
\centering
\includegraphics[width=0.45\textwidth]{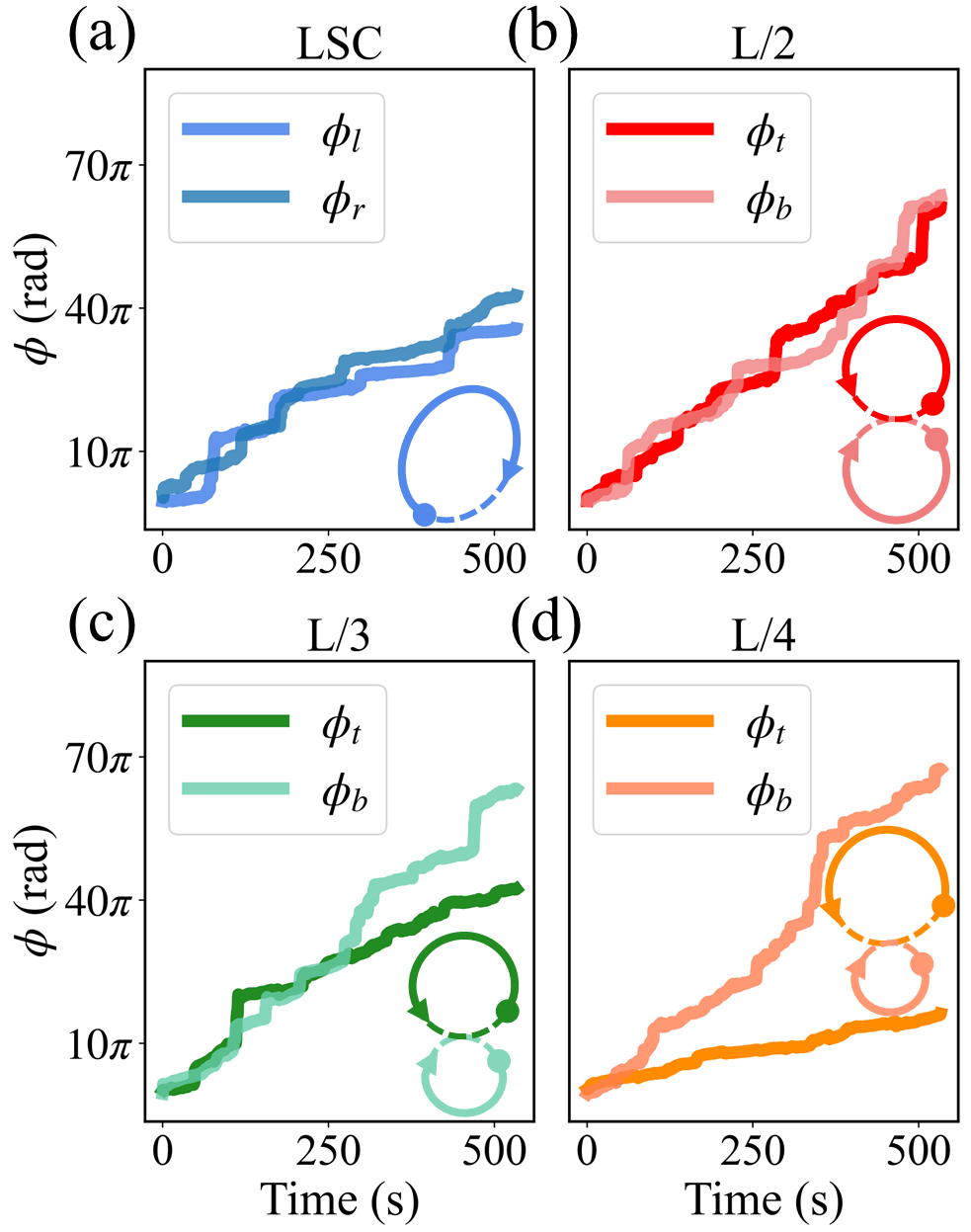}
\caption{\label{ffr}Temporal evolution of the near-wall vertical velocity signal phase $\phi(t)$ for different barrier positions at $\mathrm{Ra}=1.03 \times 10^{9}$. (a) Using the LSC as a reference, the signal phases on opposite walls evolve synchronously at the same frequency $2\pi f=\partial\phi/\partial t$, with $f_l \approx f_r = 10\;\mathrm{mHz}$ (evaluated at left $l$ and right $r$ sides). (b) When the barrier is placed at $z=L/2$, the frequency doubles ($f_t \approx f_b = 20\;\mathrm{mHz}$), while the phases remain locked. (c) With barrier at $z=L/3$, the top and bottom rolls become asymmetric ($f_t = 10\;\mathrm{mHz}, f_b = 26\;\mathrm{mHz}$) and the phases no longer evolve together. (d) With barrier at $z=L/4$, the disparity increases ($f_t = 4\;\mathrm{mHz}, f_b = 28\;\mathrm{mHz}$). 
} 
\end{figure}

To test the robustness of this relation, we create asymmetric top and botom pairs by positioning the barrier at $z = L/3$ and $L/4$ (Fig. S3, Supplemental Material), yielding structures with unequal characteristic lengths. Remarkably, the system exhibits signatures of generalized synchronization, where the state of one subsystem can be expressed as a function of the other, even if the two are not identical and have differing frequencies or phases \cite{pikovsky2001synchronization}. In this case, the relation $U_t / (l_t f_t) = U_b / (l_b f_b)$ connects the top ($t$) and bottom ($b$) rolls. In the symmetric case, the velocities and length scales associated with $t$ and $b$ are equal, constraining the system to a state where both vortices circulate at similar frequencies, approximately twice that of the original no-barrier LSC, with their phases evolving in tandem, Fig. \ref{f31}(d), Fig. \ref{ffr}(a,b). When $l_t > l_b$, asymmetries in velocity and length scales produce frequency differences, giving rise to major and minor rolls with mismatched frequencies ($f_t < f_b$) and phase evolutions, Fig. \ref{ffr}(c). The frequency mismatch and phase separation grow with the disparity between the rolls, Fig. \ref{ffr}(d). Strikingly, despite the asymmetry, the structures remain dynamically coupled (see Fig. \ref{f31}(c) and Table I in the Supplemental Material). The observed coupling likely arises because the top-bottom and left–right rolls can be viewed as branches of a single distorted, figure-eight-axisymmetric circulation, continuously exchanging information rather than forming fully separated structures \cite{xia1997turbulent}.

We also examine the system's thermal response. Flow topology and BL thickness are theorized to strongly influence heat transfer \cite{ahlers2009heat}, quantified by the Nusselt number $\mathrm{Nu} = q L / (k \Delta)$, where $q$ is the heat flux and $k$ the fluid’s thermal conductivity. Although some studies report a dependence of $\mathrm{Nu}$ on flow structure and symmetry \cite{vanderpoel2011connecting, schindler2022collapse, xu2025restoration}, others find minimal effects when the LSC is absent or forced to lose coherence \cite{ciliberto1996large, xia1997turbulent, xia1999turbulent}. Our DNS results confirm that the presence of the barrier affects $\mathrm{Nu}$ by at most 2\% relative to the no-barrier case (Fig. S2, Supplemental Material), showing that heat transport is largely unaffected, even after the flow transitions to an axisymmetric structure with thinner boundary layers.

In conclusion, we show that the hallmark structure of turbulent thermal convection, the LSC, possesses an internal `clock' $U/(lf)\approx2.1$, relating characteristic velocity $U$ to its pulsation frequency $f$ and characteristic spatial scale $l$. This dimensionless number can be physically interpreted as the imposed shear $U/l$ compared with the plume shedding frequency $f$. The LSC can survive or be tripped in the presence of a thin ring-shaped sidewall barrier in a cylindrical cell. At low thermal driving ($\mathrm{Ra}\lesssim8\times 10^8$), the LSC bypasses the barrier, sustained by thick kinetic boundary layers. At higher forcing ($\mathrm{Ra}\gtrsim8\times 10^8$), as the boundary layers thin to become comparable to the barrier scale, the flow transitions into a pair of vertically stacked, counter-rotating, interconnected vortices. We find that, while the characteristic velocity remains largely unchanged (for the case $L/2$ ), the pulsation frequency adapts to conserve $U/(lf)$ for each of the two HCCs in view of the reduced characteristic length. The rolls behave as coupled oscillators, with this relation persisting even when they become asymmetric. In this way, we uncover a precise quantity linking plume pulsation and the flow structure, showing that topology control via geometric constraints in RBC provides a platform to reveal how turbulent flows reorganize and even synchronize. %
It is of great interest to assess the general validity of these principles when moving to more complex (geophysically or industrially inspired) domains and/or other convection liquids (including multi-layer media, e.g., \cite{wang2024twolayer}).

\begin{acknowledgments}
    This work is part of the project ``Universal critical transitions in constrained turbulent flows" (file number VI.C.232.026) within the NWO-Vici research program, funded by the Dutch Research Council (NWO). The authors also thank Gerald Oerlemans, Freek van Uittert, J\o rgen van der Veen, and Chiel Koster for their technical support.
\end{acknowledgments}

\bibliography{apssamp}%

\end{document}